\documentclass[runningheads]{llncs}
\usepackage{graphicx}
\usepackage[colorlinks, linkcolor=blue, anchorcolor=blue, citecolor=blue]{hyperref}
\usepackage{marvosym}
\usepackage{cite}
\usepackage{amsmath}
\usepackage{amssymb}
\usepackage{pfnote}
\usepackage{multicol}
\usepackage{multirow}
\usepackage{booktabs}
\usepackage{float}
\usepackage{ulem}
\begin{document}
\title{CSR-dMRI: Continuous Super-Resolution of Diffusion MRI with Anatomical Structure-assisted Implicit Neural Representation Learning}
%
%
\author{Ruoyou Wu\inst{1,2,3} \and
Jian Cheng\inst{4} 
\and
Cheng Li\inst{1}
\and
Juan Zou\inst{5}
\and 
Jing Yang\inst{1,3}
\and 
Wenxin Fan\inst{1,3}
\and 
Yong Liang\inst{2}
\and
Shanshan Wang\inst{1} 
$^{\href{mailto: ss.wang@siat.ac.cn}{(\textrm{\Letter})}}$
}
\authorrunning{R.Wu et al., Continuous SR of Diffusion MRI}

\institute{Paul C. Lauterbur Research Center for Biomedical Imaging, Shenzhen Institute of Advanced Technology, Chinese Academy of Sciences, Shenzhen 518055, China 
\email{\\ ss.wang@siat.ac.cn}\\
\and Pengcheng Laboratory, Shenzhen 518055, China
\and University of Chinese Academy of Sciences, Beijing 100049, China
\and School of Computer Science and Engineering, Beihang University, Beijing 100191, China
\and School of Physics and Electronic Science, Changsha University of Science and Technology, Changsha 410114, China}

\maketitle              
\begin{abstract}
Deep learning-based dMRI super-resolution methods can effectively enhance image resolution by leveraging the learning capabilities of neural networks on large datasets. However, these methods tend to learn a fixed scale mapping between low-resolution (LR) and high-resolution (HR) images, overlooking the need for radiologists to scale the images at arbitrary resolutions. Moreover, the pixel-wise loss in the image domain tends to generate over-smoothed results, losing fine textures and edge information. To address these issues, we propose a novel continuous super-resolution method for dMRI, called CSR-dMRI, which utilizes an anatomical structure-assisted implicit neural representation learning approach. Specifically, the CSR-dMRI model consists of two components. The first is the latent feature extractor, which primarily extracts latent space feature maps from LR dMRI and anatomical images while learning structural prior information from the anatomical images. The second is the implicit function network, which utilizes voxel coordinates and latent feature vectors to generate voxel intensities at corresponding positions. Additionally, a frequency-domain-based loss is introduced to preserve the structural and texture information, further enhancing the image quality. Extensive experiments on the publicly available HCP dataset validate the effectiveness of our approach. Furthermore, our method demonstrates superior generalization capability and can be applied to arbitrary-scale super-resolution, including non-integer scale factors, expanding its applicability beyond conventional approaches.  

\end{abstract}
%
%
\section{Introduction}
Diffusion magnetic resonance imaging (dMRI) can reflect early changes in the microstructure of brain tissue in neurological diseases by measuring the diffusion displacement distribution of water molecules in the brain tissue\cite{razek2018assessment,li2022artificial}. As a non-invasive method, dMRI is extensively utilized for the diagnosis of brain diseases. Moreover, it provides a distinctive avenue for exploring the neural foundations of human cognitive behavior. However, to accurately estimate quantitative evaluation parameters for diffusion tensor imaging (DTI) or diffusion kurtosis imaging (DKI), it is typically necessary to acquire HR data, resulting in prolonged acquisition time for dMRI. This can lead to motion artifacts and patient discomfort. In clinical practice, reducing acquisition time is often achieved by sacrificing the spatial resolution of dMRI images, which can affect the estimation accuracy of DTI or DKI parameters.
Therefore, the reconstruction of HR dMRI images from LR dMRI images holds significant clinical value. 

Image post-processing methods are one of the feasible solutions to address the issue. There are some traditional interpolation methods, such as nearest-neighbor interpolation and linear interpolation, that can be utilized. However, Van et al.,\cite{van2006image} pointed out that interpolation methods often lead to blurred image edges and are unable to recover fine details. On the other hand, super-resolution reconstruction (SR) methods are an interesting alternative solution that can generate HR images from LR images\cite{wang2021deep,park2003super, umirzakova2023medical,wang2024knowledge,wang2021review}. The existing SR methods can be roughly divided into two categories: model-based methods\cite{nedjati2008regularized,scherrer2011super,wang2022parcel,tobisch2014model,ning2015compressed,shi2015lrtv,shi2016super} and data-driven methods\cite{luo2022diffusion,chen2018brain,du2020super}. Model-based methods rely on mathematical models to build connections between LR and HR images, while data-driven methods utilize neural networks to learn the nonlinear mapping relationship between LR and HR images. With the fast development of deep learning (DL), SR methods based on deep learning have been widely investigated\cite{wang2020deep,lim2017EDSR,zhang2018RDN}. For example, Lim et al.,\cite{lim2017EDSR} proposed EDSR, which improves super-resolution image reconstruction quality by stacking deeper or wider networks to extract more information within the same computational resources. Zhang et al.,\cite{zhang2018RDN} introduced the RDN network, which utilizes residual dense blocks with densely connected convolutional layers to extract rich local features. It also allows direct connections from all layers of the previous residual dense block (RDB) to the current RDB, making full use of hierarchical features extracted from the original LR image. These methods have shown promising performance in enhancing the quality of the reconstructed super-resolution images. 

Recently, some researchers have started applying SR techniques to diffusion MRI\cite{luo2022diffusion,chatterjee2021shuffleunet}, primarily using learning-based methods. For instance, Chatterjee et al.,\cite{chatterjee2021shuffleunet} proposed a ShuffleUNet architecture to reconstruct super-resolution dMRI data, addressing issues of image blurring and over-smoothing by replacing stridden convolutions with lossless pooling layers. Luo et al.,\cite{luo2022diffusion} proposed a sub-pixel convolution generative adversarial attentional network (SPC-GAAN) for the reconstruction super-resolution dMRI data, achieving promising results. However, these methods can only reconstruct super-resolution dMRI data with a fixed integer scale factor, limiting their applicability. In practical applications, due to differences in scanning protocols, the resolution of acquired dMRI data is inconsistent. To better facilitate clinical analysis, it is necessary to reconstruct them to a consistent resolution. The aforementioned methods may face challenges in achieving this. Additionally, the pixel-wise loss in the image domain, which was utilized in these methods, tends to generate over-smooth results, leading to the loss of fine textures and edge information.

To address these issues, we propose an anatomical structure-assisted implicit neural representation learning framework for continuous super-resolution of dMRI images. To the best of our knowledge, this is the first attempt of implicit neural representation learning in dMRI super-resolution. Specifically, we introduce implicit neural representation into the super-resolution of dMRI, enabling arbitrary-scale super-resolution for dMRI, and it can be used for transforming dMRI data at different resolutions. Our contributions can be summarized as follows:

1) We propose a novel paradigm for arbitrary-scale super-resolution in diffusion MRI by combining anatomical structure-assisted implicit neural representation learning, called CSR-dMRI. It can be used to transform dMRI data at different resolutions into dMRI data at the same resolution, facilitating clinical analysis. 

2) The details and texture information of the image are improved by introducing anatomical image and frequency-domain-based loss.

3) Extensive experiments on the public HCP dataset demonstrate the effectiveness of our approach. Furthermore, our method exhibits better generalization and can be applied to non-integer scale factors, expanding the applicability of this approach.

\section{Methods}
\subsection{Implicit Neural Representation Learning}
In implicit neural representation learning, a voxel can be represented by a neural network as a continuous function. the network $\mathcal{F}_{\theta}$ with parameters $\theta$ can be defined as:

\begin{equation}
    \label{eq1}
    I=\mathcal{F}_{\theta}(c),\ c\in [-1, 1]^{3},\ I\in \mathbb{R}^{3}   
\end{equation}
where the input $c$ is the normalized coordinate index in the voxel spatial field, and the output $I$ is the corresponding intensity value in the voxel. The network function $\mathcal{F}$ maps coordinates to voxel intensities, effectively encoding the internal information of the entire voxel into the network parameters. The network $\mathcal{F}$ with parameters $\theta$ is also referred to as the neural representation of the voxel. Our voxel coordinates are constructed based on the size of the dMRI data and normalized along each dimension to the range of [-1,1]. This network is only applicable to the super-resolution of individual dMRI images and is not suitable for multiple dMRI images. Additional auxiliary information needs to be provided.

\begin{figure}
\centering
\includegraphics[width=0.9\textwidth]{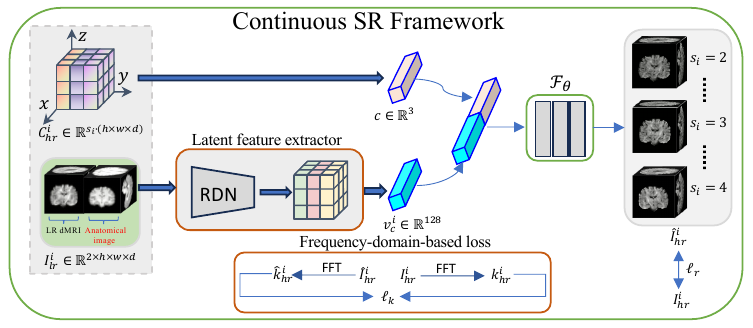}
\caption{The overall structure of the proposed CSR-dMRI model. The latent feature extractor is used to extract latent space feature maps from LR dMRI and anatomical images. Implicit function network $\mathcal{F}_{\theta }$ is used to predict voxel intensity at spatial coordinate $c$. Frequency-domain-based loss is used to improve the details and texture information of the image.} \label{fig1}
\end{figure}

\subsection{Model Overview}
The overall architecture of the proposed CSR-dMRI model is illustrated in Figure \ref{fig1}. Due to the primary focus of dMRI data on tissue microstructure and diffusion information, which provides limited anatomical information. So, we introduce T1-weighted data to provide anatomical prior information. Additionally, by incorporating frequency-domain-based loss to preserve texture and structural information in the image, the quality of the image is enhanced. The entire network consists of a latent feature extractor and an implicit function network. The latent feature extractor is used to extract features from LR dMRI and anatomical images, while the implicit function network predicts voxel intensities at arbitrary coordinates by integrating coordinates and corresponding latent feature vectors, thereby achieving super-resolution. Specifically, for a given pair of LR dMRI and anatomical images from the dataset $D=\{{I_{lr}^{i}\in \mathbb{R}^{2\times h\times w\times d}, I_{hr}^{i}\in \mathbb{R}^{s_{i}\cdot (h\times w\times d) }    } \}_{i=1}^{N}$, along with their corresponding HR dMRI, where $N$ represents the total number of samples, and $s_{i}$ represents the upsampling scale factor for the $i$-th sample pair, dimension 2 indicates concatenating DWI images and anatomical images along the channel. We first utilize a latent feature extractor to transform the LR dMRI and anatomical images into feature maps $v_{lr}^{i}\in \mathbb{R}^{h\times w\times d\times 128} $, where each element $v_{c}^{i}\in \mathbb{R}^{128}$ corresponding to the feature vector of the LR dMRI at coordinate $c$. For any voxel coordinate $c$ in HR dMRI, we generate the corresponding feature map $v_{hr}^{i}\in\mathbb{R}^{s_{i}\cdot(h\times w\times d)}$ by bilinear interpolation of $v_{lr}^{i}$. Subsequently, the queried coordinate $c$ and the corresponding feature vector $v_{c}^{i}$ are concatenated as input to $\mathcal{F}_{\theta }$, and the output of the $\mathcal{F}_{\theta }$ is the predicted voxel intensity $\hat{I}^{i}_{hr}(c) $ at spatial coordinate $c$. By minimizing the difference between the predicted voxel intensity $\hat{I}^{i}_{hr}(c) $ and the real voxel intensity $I^{i}_{hr}(c) $, both the latent feature extractor and implicit function network are simultaneously optimized.

\subsubsection{Latent Feature Extractor.} For the latent feature extractor, inspired by the works\cite{chen2021learning,wu2022arbitrary}, Residual Dense Network (RDN)\cite{zhang2018RDN} is used to extract latent space feature vectors from LR dMRI and anatomical image data. The latent feature extractor takes the LR data $I_{lr}^{i}\in \mathbb{R}^{2\times h \times w\times d}$ as input and outputs the feature map $v_{lr}^{i}\in\mathbb{R}^{h\times d\times w\times128}$. This latent space feature extraction approach helps the implicit function network $\mathcal{F}_{\theta}$ effectively integrate local information of the image, enabling the recovery of details in HR images even at large upsampling scale factors.
Additionally, to maintain the scale of the extracted features uncharged, we removed the upsampling operation from the last layer of the original RDN network and expanded all 2D convolutional layers to 3D convolutional layers. To ensure extracting a sufficient amount of features for each coordinate position, the output channel number of the last layer is set to 128.

\subsubsection{Implicit Function Network.} The implicit function network $\mathcal{F}_{\theta}$ consists of a sequence of 8 consecutive fully connected layers, each followed by a $ReLU$ activation layer. A residual connection is established between the input of the network and the output of the fourth $ReLU$ activation layer. The goal of $\mathcal{F}_{\theta}$ is to predict the voxel intensity at any spatial coordinate $c$. The specific process is as follows:

\begin{equation}
\hat{I}^{i}_{hr}=\mathcal{F}_{\theta }(c, v_{c}^{i}) 
\label{eq2}
\end{equation}
where $c$ represents the spatial coordinate position to be predicted, $v_{c}^{i}$ represents the specific latent feature vector of the voxel at spatial coordinate $c$. Instead of using spatial coordinates alone as inputs to the implicit function network, we combine the spatial coordinates with their corresponding latent feature vectors. This approach effectively integrates the local semantic information from the image, thereby enhancing the ability of $\mathcal{F}_{\theta}$ to recover image details. 

\subsubsection{Loss Function.} In image super-resolution tasks, the pixel-wise loss is commonly used as a loss function to improve the model's performance and convergence\cite{zhao2016loss}. However, since pixel-wise loss does not take into account the perceptual quality of the image, the generated results tend to be smooth and lack high-frequency details. To address this issue, we introduce a frequency-domain-based loss that better captures the frequency information of the image, preserving both the structure and texture details in the image, thereby further improving the image quality. The specific process is as follows:

\begin{equation}
\ell _{k}=\frac{1}{N}\sum_{i=1}^{N}\left | \hat{k}_{hr}^{i} - k_{hr}^{i}  \right |  
    \label{eq3}
\end{equation}
where $\hat{k}_{hr}^{i}$ and $k_{hr}^{i}$ represent the fast Fourier transforms of $\hat{I}_{hr}^{i}$ and $I_{hr}^{i}$, respectively, as shown in Figure \ref{fig1}. $N$ denotes the total number of samples. Finally, the objective function of the CSR-dMRI model is defined as:

\begin{equation}
\ell _{f}=\ell_{r}+\lambda _{k}\ell _{k}
    \label{eq4}
\end{equation}
we set $\lambda_{k}$=0.01 empirically to balance the two losses, $\ell_{r}$ denotes the L1 loss between $\hat{I}_{hr}^{i}$ and $I_{hr}^{i}$. 

\section{Experiments}
\subsection{Datasets}
In this paper, we utilize data from the Human Connectome Project (HCP) WU-Minn-Ox Consortium public dataset\footnote{https://www.humanconnectome.org}. We select 100 pre-processed diffusion and T1-weighted MRI data. Out of these, 70 were used for training, 10 for validation, and 20 for testing. The whole-brain diffusion MRI data were acquired with an isotropic resolution of 1.25mm, featuring four b-values (0, 1000, 2000, 3000). For diffusion MRI data, we only utilize the data from b1000, normalized by dividing the b1000 data by the average value of 18 b0 data. Firstly, we extract 9 patches of $40^{3}$ dimension size from each sample. Subsequently, these patches were cropped to generate HR patches of $(10\times s)^3$ dimension size. Finally, the HR patches were downsampled to obtain LR patches of $10^3$ dimension size using bicubic interpolation. The scale $s$ is randomly sampled from a uniform distribution $\mathcal{U}(2,3) $. 

\subsection{Experimental Setup}
All networks are trained using the Pytorch framework with one NVIDIA RTX A6000 GPU (with 48 GB memory). The Adam optimizer is employed for model training with an initial learning rate of $10^{-4}$, and the batch size is set to 9. Every 200 epochs, the learning rate is multiplied by 0.5. The model is trained for a total of 1000 epochs.
We compare our CSR-dMRI model with some state-of-the-art methods, including traditional interpolation method: Bicubic, deep learning-based fixed-scale super-resolution methods: DCSRN\cite{chen2018brain}, ResCNN3D\cite{du2020super}, and an arbitrary scale super-resolution method: ArSSR\cite{wu2022arbitrary}. PSNR and SSIM are used for quantitative evaluation. 

\subsection{Results}
To comprehensively assess the effectiveness of our approach, we conduct experiments in three aspects: the first involves in-distribution experiments, including $2\times$ and $3\times$ experiments.The second is out-of-distribution experiments where we tested the results under the $4\times$. The third involves experiments with non-integer scale factors, testing the results under the $2.4\times$. 

\begin{table}[ht]
\centering
\caption{Quantitative results of different methods on the HCP dataset. Bold numbers indicate the best results. }
\label{tab1}
\renewcommand{\arraystretch}{1.5} 
\resizebox{0.9\textwidth}{!}{
\tiny 
\begin{tabular}{ccccccccc}
\hline
                         & \multicolumn{4}{c}{In-distribution}                                                                                                                                     & \multicolumn{2}{c}{Out-of-distribution}                                            & \multicolumn{2}{c}{Non-integer scale}                                              \\ \cline{2-9} 
                         & \multicolumn{2}{c}{2$\times$}                                                             & \multicolumn{2}{c}{3$\times$}                                                             & \multicolumn{2}{c}{4$\times$}                                                             & \multicolumn{2}{c}{2.4$\times$}                                                           \\ \cline{2-9} 
\multirow{-3}{*}{method} & PSNR                                     & SSIM                                    & PSNR                                     & SSIM                                    & PSNR                                     & SSIM                                    & PSNR                                     & SSIM                                    \\ \hline
Bicubic                  & 25.8880          & 0.9075          & 23.6657          & 0.8360          & 22.6657          & 0.7774          & 24.6268          & 0.8778          \\
DCSRN                    & 26.4039          & 0.8901          & 24.3236          & 0.8479          & 23.1425          & 0.7974          & 25.1044          & 0.8797          \\
ResCNN3D                 & 26.7900          & 0.9291          & 24.7841          & 0.8667          & 23.3626          & 0.8135          & 25.4493          & 0.8927          \\
ArSSR                    & 26.6123          & 0.9343          & 25.3040          & 0.9059          & 23.8770          & 0.8554          & 26.2907          & 0.9266          \\ 
\textbf{CSR-dMRI}            & \textbf{27.3611} & \textbf{0.9458} & \textbf{26.0762} & \textbf{0.9235} & \textbf{24.5061} & \textbf{0.8752} & \textbf{27.1196} & \textbf{0.9410} \\ \hline
\end{tabular}
}
\end{table}

\subsubsection{Quantitative Results.} Table \ref{tab1} shows the quantitative results of different methods under in-distribution, out-of-distribution, and non-integer scale factor conditions. Since the training scales for the DCSRN and ResCNN3D models are fixed at $2\times$ and $3\times$, respectively, when testing with the non-integer scale factor of $2.4\times$, we first use the $3\times$ model for super-resolution and then perform downsampling to obtain the $2.4\times$ results. For testing with the out-of-distribution scale factor of $4\times$, we first use the $3\times$ model for super-resolution and then upsampling to obtain the $4\times$ results. The arbitrary-scale super-resolution method ArSSR and our CSR-dMRI model are both trained at scale factors of $2\times$ to $3\times$. Overall, our method achieves the best results across all scales. In out-of-distribution scenarios, our method outperforms existing approaches and maintains better performance even with non-integer scale factors. The experimental results confirm that our method excels in performance and generalization compared to existing methods, and it can also accommodate non-integer scale factors effectively, meeting the diverse imaging resolution needs of medical professionals.

\begin{figure}[ht]
\centering
\includegraphics[width=0.9\textwidth]{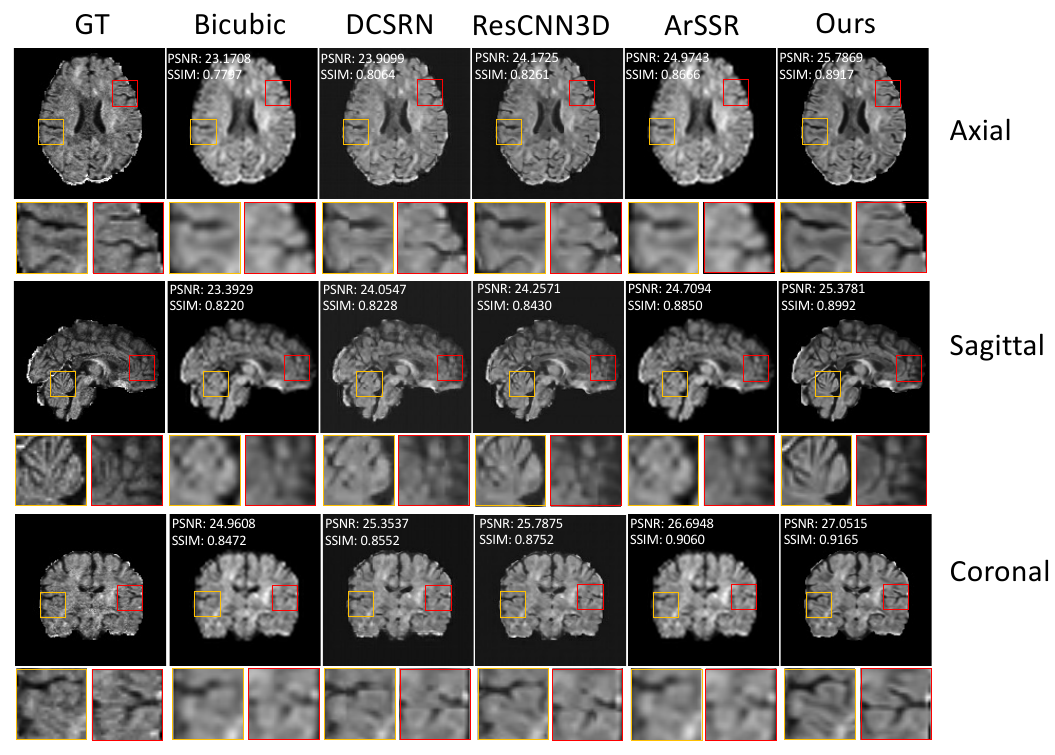}
\caption{Qualitative results of different methods on the HCP dataset and corresponding local zoomed-in results. Including the axial, sagittal, and coronal directions. The corresponding quantitative metrics are presented in the top-left corner.} \label{fig2}
\end{figure}

\subsubsection{Qualitative Results.} To further compare the performance of the models, Figure \ref{fig2} displays the reconstruction results of various methods at a non-integer scale ($2.4\times$) in the axial, sagittal, and coronal directions, along with corresponding local zoomed-in results for two areas. From a visual perspective, our method can preserve more detailed information. Compared to ArSSR, our method CSR-dMRI performs better in terms of image details, benefiting from the introduced anatomical images and frequency-domain-based loss constraints. Additionally, locally magnified images also indicate that our method can eliminate block artifacts and preserve better texture details. Overall, our method achieves the best results.

\subsection{Ablation Study}
To evaluate the effectiveness of key components in the CSR-dMRI model, we conduct ablation experiments on different components, as shown in Table \ref{tab2}. INR indicates whether implicit neural representations are used, as ArSSR\cite{wu2022arbitrary}, T1w indicates the inclusion of anatomical image, $\ell_{k}$ indicates the introduction of frequency-domain-based loss. The quantitative results in Table \ref{tab2} indicate that different components have positively contributed to the results. Furthermore, the structural prior information provided by anatomical images effectively enhances the image quality.

\begin{table}[ht]
\centering
\caption{Ablation experiments of different components on the HCP dataset. Bold numbers indicate the best results. }
\label{tab2}
\renewcommand{\arraystretch}{1.5} 
\resizebox{0.9\textwidth}{!}{
\tiny 
\begin{tabular}{cccccccccccc}
\hline
                         &                       &                                                                               &                           & \multicolumn{4}{c}{In-distribution}                                                                                                                                     & \multicolumn{2}{c}{out-of-distribution}                                            & \multicolumn{2}{c}{Non-integer scale}                                              \\ \cline{5-12} 
                         &                       &                                                                               &                           & \multicolumn{2}{c}{2x}                                                             & \multicolumn{2}{c}{3x}                                                             & \multicolumn{2}{c}{4x}                                                             & \multicolumn{2}{c}{2.4x}                                                           \\ \cline{5-12} 
\multirow{-3}{*}{Setting} & \multirow{-3}{*}{INR} & \multirow{-3}{*}{T1w} & \multirow{-3}{*}{$\ell_{k}$}   & PSNR                                     & SSIM                                    & PSNR                                     & SSIM                                    & PSNR                                     & SSIM                                    & PSNR                                     & SSIM                                    \\ \hline
Baseline                 & $\times$                     & $\times$                                                     & $\times$ & 26.9584          & 0.9352          & 24.8094          & 0.8789          & 23.3321          & 0.8240          & 25.4377          & 0.9015          \\
M1                       & $\surd$                     & $\times$                                                                             & $\times$                         & 26.6123          & 0.9343          & 25.3040          & 0.9059          & 23.8770          & 0.8554          & 26.2907          & 0.9266          \\
M2                       & $\times$                     & $\surd$                                                                             & $\times$                         & 27.0454          & 0.9444          & 25.8534          & 0.9222          & 24.2769          & 0.8734          & 26.9627          & 0.9408          \\
M3                       & $\times$                     & $\times$                                                                             & $\surd$                         & 26.8047          & 0.9354          & 25.5623          & 0.9089          & 24.0327          & 0.8575          & 26.5996          & 0.9294          \\ 
\textbf{CSR-dMRI}          & \textbf{$\surd$}            & \textbf{$\surd$}                                                                    & \textbf{$\surd$}                & \textbf{27.3611} & \textbf{0.9458} & \textbf{26.0762} & \textbf{0.9235} & \textbf{24.5061} & \textbf{0.8752} & \textbf{27.1196} & \textbf{0.9410} \\ \hline
\end{tabular}
}
\end{table}

\vspace{-0.6cm}
\section{Conclusion}
In this paper, we propose a CSR-dMRI method for continuous super-resolution of diffusion MRI with anatomical structure-assisted implicit neural representation learning. Details and texture information in the image are preserved by incorporating anatomical images and frequency-domain-based loss during training. Extensive experiments on the HCP dataset indicate superior performance and generalization of our CSR-dMRI model, showing applicability across non-integer scale factors. This contributes to addressing the clinical demand for images at different resolutions.

\subsubsection{Acknowledgements.}
This research was partly supported by the National Natural Science Foundation of China (62222118, U22A2040), Guangdong Provincial Key Laboratory of Artificial Intelligence in Medical Image Analysis and Application (2022B1212010011), Shenzhen Science and Technology Program (RCYX20210706092104034, JCYJ20220531100213029), and Youth Innovation Promotion Association CAS.

%
%
%

\bibliographystyle{splncs04}
\normalem
\bibliography{ref}
%




\end{document}